\documentclass[a4paper,11pt]{article}
\pdfoutput=1 

\usepackage{jcappub} 

\usepackage[T1]{fontenc} 
\usepackage[utf8]{inputenc}
\usepackage[colorlinks=true]{hyperref}
\usepackage{graphicx,bm,amsmath,color}
\usepackage{bbold}
\usepackage{wasysym}
\usepackage{verbatim}
\usepackage{amssymb}
\usepackage{epstopdf}
\usepackage{animate}
\usepackage{xmpmulti}
\usepackage{centernot}
\usepackage{multirow}
\usepackage{tikz}
\usepackage{graphicx}
\usepackage{float}
\usepackage{mathtools}
\usepackage{comment}

\newcommand{\be}{\begin{equation}}
\newcommand{\ee}{\end{equation}}
\newcommand{\beq}{\begin{eqnarray}}
\newcommand{\eeq}{\end{eqnarray}}
\newcommand{\ba}{\begin{align}}
\newcommand{\ea}{\end{align}}

\title{Exploring black holes as particle accelerators: hoop-radius, target particles and escaping conditions}

\author[a,b]{Stefano Liberati}
\author[c]{Christian Pfeifer}
\author[d,e,f,1]{and Javier Relancio,\note{Corresponding author.}}

\affiliation[a]{SISSA\\ Via Bonomea 265, 34136 Trieste, Italy and INFN, Sezione di Trieste}
\affiliation[b]{IFPU - Institute for Fundamental Physics of the Universe\\ Via Beirut 2, 34014 Trieste, Italy}
\affiliation[c]{ZARM, University of Bremen,\\28359 Bremen, Germany}
\affiliation[d]{Dipartimento di Fisica ``Ettore Pancini'', Università di Napoli Federico II, 80138 Napoli, Italy}
\affiliation[e]{INFN, Sezione di Napoli, 80126 Napoli, Italy;}
\affiliation[f]{Centro de Astropartículas y Física de Altas Energías (CAPA), Universidad de Zaragoza,\\Zaragoza 50009, Spain}

\emailAdd{liberati@sissa.it}
\emailAdd{christian.pfeifer@zarm.uni-bremen.de}
\emailAdd{relancio@unizar.es}

\abstract{The possibility that rotating black holes could be natural particle accelerators has been subject of intense debate. While it appears that for extremal Kerr black holes arbitrarily high center of mass energies could be achieved, several works pointed out that both theoretical as well as astrophysical arguments would severely dampen the attainable energies. In this work we study particle collisions near Kerr black holes, by reviewing and extending the so far proposed scenarios. Most noticeably, we shall focus on the recently advanced target particle scenarios which were claimed to reach arbitrarily high energies even for Schwarzschild black holes. 
By implementing the hoop conjecture we show that these scenarios involving near-horizon target particles are in principle able to attain, sub-Planckian, but still ultra-high center of mass energies of the order of $10^{23}-10^{25}$ eV even for non-extremal Kerr black holes. Furthermore, analysing the properties of particles produced in such collisions, we find that photons can escape to infinity. However, their energy is only of the order of the energy of the colliding particles and hence relatively low, which is the same  conclusion previously reached in the literature about  the original Bañados--Silk--West process. This finding points towards a general limitation of collisional Penrose processes, at least for what concerns their primary products.
}

\begin{document}
\maketitle
\flushbottom

\section{Introduction}
Since Penrose's original paper~\cite{Penrose:1971uk}, pointing out the possibility to exploit rotating black holes' ergoregions to extract energy, there have been several efforts in the literature aiming at developing and optimizing this idea. In~\cite{Banados:2009pr}, Bañados--Silk--West (BSW) observed that collisions of particles with specific angular momentum values produce arbitrarily high center of mass energies in extremal Kerr black hole spacetimes. This collisional Penrose process has been studied for different black hole scenarios, such as non-extremal Kerr~\cite{Grib:2010dz,Lake:2010bq,Gao:2011sv,Ogasawara:2016yfk}, Kerr--(anti)de Sitter~\cite{Li:2010ej,Zhang:2018ocv}, Reissner--Nordstr\"om~\cite{Zaslavskii:2010aw,Nakao:2017xwe}, and Kerr~\cite{Wei:2010vca,Liu:2011wv,Sadeghi:2020flf} geometries, reproducing the original BSW result. Also, taking into account the spin or the charge of the particles, one still obtains an arbitrarily high center of mass energy for collisions near to the horizon of rotating black holes \cite{Zhang:2018gpn, Mukherjee:2018kju,Maeda:2018hfi,Jiang:2019cuc}. For non-rotating black holes this feature is absent \cite{Armaza:2015eha}, as long as the colliding particles fall towards the black hole from large (infinite) radial distance. 

An alternative mechanism, leading to infinite center of mass energies for collisions in non-rotating black hole spacetimes, has been proposed in \cite{Grib:2011ph}, and more recently been brought forward again in~\cite{Hackmann:2020ogy}. The difference with respect to the BSW setup is that only one of the colliding particles comes from infinity while the other particle (the target particle) is created (or placed) close to the horizon. The latter feature means that the energy parameter of the particle trajectory of this second particle is small, which can be used to show that even in a non-rotating black hole spacetime, arbitrarily large center of mass energies can be achieved~\cite{Grib:2011ph,Hackmann:2020ogy}. Following the same idea, a similar result can be derived in the case in which the spin of the particles is also taken into account~\cite{Sheoran:2020kmn}. For rotating black holes, it is even possible to consider the scenario of the collision of one infalling particle coming from infinity and a zero or negative energy or zero angular momentum particle, propagating inside the ergosphere~\cite{Grib:2016mak,Zaslavskii:2016uhv}. As discussed in~\cite{Wald:1974kya}, particles with negative energy can be produced in the  ergosphere of a black hole by the decay of particles coming from infinity. Depending on the particular values of the parameters (velocity and angular momentum), zero energy particles can also be emitted. But this Penrose's process is not the only way in which particles with zero or almost zero energy can be produced. For example, a collision inside the ergoregion can lead to this kind of particles.   

Coming back to the  above described collisional Penrose process, a common objection is that it is derived in an over idealized situation. Indeed, in the extant literature many arguments have been brought forward about why this infinite center of mass energy cannot emerge in a realistic physical situation. In particular by pointing out that in realistic situations several limiting mechanisms would not only prevent these unbounded energies but also lead to relatively low values.

The first of such mechanisms worth mentioning is the so called ``hoop conjecture''. Proposed by Thorne in 1972~\cite{Klauder:1972lsv}, it basically states that the presence of a particle with non-negligible energy in the vicinity of a black hole induces a backreaction which leads to a new, larger, horizon before the particle crosses the original one. This conjecture has been specialized to static, vacuum, axisymmetric spacetimes, and for oblate geometries~\cite{PhysRevD.44.2409}, and also in the spatially regular static charged fluid sphere spacetimes~\cite{Peng:2019cal}. This feature (in fact a weaker version of this statement~\cite{Hod:2015iig}), was used in~\cite{Hod:2016kzj} to show that the above mentioned collisions cannot take place as close to the black hole horizon as needed for achieving ultra-high energies, but always at a greater radius limited below by the hoop conjecture.  The implementation of the hoop radius, which would be the new event horizon of the enlarged black hole if the particles would collide beyond it, avoids the prediction of infinite energy collisions outside an event horizon~\cite{Hod:2016kzj}. More recently, in~\cite{Hod:2020eau}, it was provided evidence for the non-existence of a unified version of the hoop conjecture which is valid for both black-hole spacetimes and spatially regular horizonless compact objects.

Other important limiting factors which one needs to take into account when searching for such high center of mass collisions near black holes are: the Thorne limit~\cite{Thorne:1974ve}, which states that the spin of realistic astrophysical black holes is bounded from above~\cite{Berti:2009bk,Jacobson:2009zg,Gao:2011sv} (but see the related discussion later on about possible violations of this limit \cite{Kesden:2008ga,Sadowski:2011ka}); the time which passes for an observer far away from the black hole, while the collision of two particles reaching the near horizon region of a black hole take place (this can be of order of the age of the universe)~\cite{Patil:2015xha,Armaza:2015eha}; multiple scatterings and collisions of the particle(s) emerging after the collision near the black hole~\cite{Leiderschneider:2015kwa,Zaslavskii:2015fqy,Zaslavskii:2019pdc}; and the question of the existence of target particles near the horizon of black holes \cite{Zaslavskii:2020agl}. All of these phenomena need to be taken into account and are basically screening the existence of a collisional Penrose process or damping its effectiveness as a particle accelerator.

In this work we begin by studying the most general scenario of a rotating black hole in Sec.~\ref{sec:kerr_newman}, with the aim to clarify the general workings of the BSW mechanism and its variants. After that, in Sec.~\ref{sec:collisional_scenarios} we apply the hoop conjecture to the three alternative scenarios: collision of two particles coming both from infinity,  one from infinity and one target particle with small energy, and one from infinity and one target particle with zero energy and small angular momentum. This last scenario has not, to our knowledge, been considered in the literature. We shall see that in all these scenarios the energy of the center of mass always stays finite (due to the hoop conjecture) but still can be very large, in particular this happens already for non-extremal Kerr black holes. 

In Sec.~\ref{sec:stellarBHs} we show that the same  energy-limiting arguments considered for the BSW mechanism are still effective in the aforementioned new scenarios. Nonetheless, we find that in the new class of collisional scenarios (those with a target particle placed closed to the hoop horizon), very high center of mass energies are generically still achievable. Indeed, the order of these energies is higher than the highest energies detected in ultra high energy cosmic rays (UHECR). We further analyse under which conditions particles produced in a 2-2 process can escape to infinity and what their energy will be. It turns out that that for nearly (but not quite) extremal black holes photons produced at the hoop radius can still escape, however their energy is  of the same order of the energies of the incoming particle, which agrees with  other results in the literature for the BSW scenario~\cite{Bejger:2012yb,Leiderschneider:2015kwa}. This means that the collisional scenario proposed in~\cite{Grib:2011ph,Hackmann:2020ogy}, and the one we studied here for the first time, are neither able to produce arbitrarily high center of mass energies nor very-high energy particles after the collision, once the hoop radius is taken into account.

This seems to suggest that, at least in principle, the considered collisions are not able to produce per se a flux of ultra high energy particles at infinity. Nonetheless, the presence of an environment containing very high energy particles close to the horizon, leaves open the issue of the possible astrophysical relevance of the expected high energy secondary products. We shall comment further about this in our conclusions, Sec.~\ref{sec:conclusions}. 

\section{Particle motion in Kerr black hole geometry}
\label{sec:kerr_newman}
In this section, we recall how the center of mass energy of particle collisions in the vicinity of a rotating Kerr black hole is derived. The metric defining this geometry is \cite{Wei:2010vca}
\begin{equation}
\begin{split}
 g_{tt}\,=&\,-\left(1-\frac{2M r}{\Sigma}\right)\,,\qquad  g_{t\varphi}\,=\,-\bar{a} \sin^2{\theta}\frac{2Mr}{\Sigma}\,,\qquad  g_{rr}\,=\,\frac{\Sigma}{\Delta}\,,\\
& g_{\theta\theta}\,=\,\Sigma\,,\qquad g_{\varphi\varphi}\,=\,\frac{\sin^2{\theta}}{\Sigma}\left( (\bar a^2 + r^2)^2 - \bar a^2 \Delta \sin^2{\theta}\right)\,,
\end{split}
\label{eq:metric_kerr_newman}
\end{equation}
where $\Sigma=r^2+\bar{a}^2\cos^2 \theta$, $\Delta=r^2+\bar{a}^2-2rM$, and $\bar{a}=J/M$, with $M$, and $J$, being the mass, and angular momentum the black hole. The horizon is characterized by the condition $\Delta = 0$, while $g_{tt}=0$ defines the so-called ergosphere.

The geodesic motion of point particles in a Kerr spacetime can be obtained by solving the Hamilton equations of motion of the Hamilton function
\begin{equation}
    H(x,k)\,=\, g^{\mu \nu} k_\mu k_\nu\,,
    \label{eq:cas}
\end{equation}
which defines the dispersion relation $H(x,k)=-m^2$ of massive ($m\neq0$) and massless ($m=0$) point particles.
We will use this  Hamiltonian approach to describe the particle motion in order to simplify the computations. 

For the purpose of this article it suffices to restrict the particle trajectories to the equatorial plane, i.e.\  $\theta = \pi/2$. Due to the static and axial symmetry of the metric Kerr geometry we can characterize the trajectories of point particles in terms of the radial momentum $k_r$. Using the dispersion relation, we can express the radial momentum as a function of the mass $m$ and the constants of motion, i.e.~the energy $E = k_t$, and the angular momentum $L = -k_\phi$
\begin{align}
   k_r 
   &\,=\,\frac{1}{\Delta}\sqrt{( E ( \bar a^2 + r^2 ) - \bar a L)^2 - ( m^2 r^2 + (\bar a E - L)^2)\Delta}\,.
    \label{eq:radial_momentum}
\end{align}

Consider two uncharged particles of equal mass $m_1=m_2=m$ with four-momenta $k_1{}_{\nu} = (E_1, k_{1r}, 0, L_1)$ and $k_2{}_{\nu} = (E_2, k_{2r}, 0, L_2)$  respectively. The squared center of mass energy at the point of collision as function of $r$ is defined as  
\begin{equation}
   E^2_{\text{cm}}(r)\,=\,-g_{\mu \nu}  \left( k^\mu_1+k_2^\mu\right)\left(k^\nu_1+k_2^\nu\right)\,.
   \label{eq:cm}
\end{equation} 
Using Eqs.~\eqref{eq:radial_momentum} and \eqref{eq:cm} we find
\begin{equation}
\begin{split}
E_{\textrm{cm}}^2(r) 
\,=\,& \frac{2}{r^2 \Delta} \Big( (E_1 (\bar a^2 + r^2 ) - \bar a L_1) (E_2 ( \bar a^2 + r^2) - \bar a L_2 )  + \Delta ( m^2 r^2 - (\bar a E_1 - L_1 ) (\bar a E_2 - L_2 ) \\
&-  \sqrt{( E_1 ( \bar a^2 + r^2 ) - \bar a L_1 )^2 - \Delta  ( ( \bar a E_1 - L_1)^2 + m^2 r^2)} \\
&
\sqrt{( E_2 ( \bar a^2 + r^2 ) - \bar a L_2 )^2 - \Delta  ( ( \bar a E_2 - L_2)^2 + m^2 r^2 )} \Big)\,.
\end{split}
\label{eq:cm_kerr_newman}
\end{equation}

For collisions at the outer black hole horizon $r_\text{horizon}=M(1+\tau)$, with 
\begin{align}\label{eq:tau}
    \tau \,=\,\sqrt{1-\left(\frac{\bar a}{M}\right)^2}\,,
\end{align}
one obtains, by taking the limit $E_{\textrm{cm}}^2(r_\text{horizon}) = \lim_{\epsilon\to0} (E_{\textrm{cm}}^2(r_\text{horizon}+\epsilon))$,
\begin{align}\label{eq:COMHor}
     E_{\textrm{cm}}^2(r_\text{horizon}) \,
     &=\,  \frac{\left(\bar a \left(m^2-E_1^2\right)+E_1 L_1\right) (E_2 L_1-E_1 L_2)}{E_1 \left(\bar a^2 E_1-\bar a L_1+E_1 M^2 (\tau +1)^2\right)}+\frac{\left(\bar a \left(m^2-E_2^2\right)+E_2 L_2\right)(E_1 L_2-E_2 L_1)}{E_2 \left(\bar a^2 E_2-\bar a L_2+E_2 M^2 (\tau +1)^2\right)}\nonumber\\
     &+\frac{m^2 (E_1+E_2)^2}{E_1 E_2}\,.
\end{align}
There are several scenarios for which this expression may become infinitely large. 

As long as $\bar a\neq0$, i.e. for rotating black holes, there exists a critical value of the angular momentum of the particles such that, if one of the colliding particles assumes this angular momentum, the denominator of one of the first two terms vanishes, as has been discussed in the literature \cite{Banados:2009pr}. A second possibility is that the energy of one of the particles is very small but non-vanishing and that its angular momentum is zero, which can be seen by applying a power series expansion for small $E_1$ or $E_2$ to \eqref{eq:COMHor}. For non-rotating black holes, this applies without any constraint on the angular momentum of the particle, as has been pointed out in \cite{Grib:2011ph,Hackmann:2020ogy}. For rotating black holes, there exists the possibility of particle trajectories in the ergosphere with vanishing energy parameter, which would lead to a divergence in the last term of \eqref{eq:COMHor}. This very last scenario has not yet been studied in the literature and we investigate it here for the first time in detail.

So far, the particles propagating through the black hole spacetime have been treated as test particles without backreaction onto the geometry. However, in case the energy of the individual particles or their center of mass energy becomes large, their backreaction cannot be neglected. One way to capture the influence of the particles on the geometry is the \emph{hoop conjecture} \cite{Klauder:1972lsv}. This conjecture states that if the compactness of an object along some closed path around it is equal to that of a black hole, then a horizon will form. In our setting, this is tantamount to say that the collision cannot happen arbitrarily close to the event horizon, but rather at a minimal radius below which the compactness would be so great that a horizon would form and entrap the particles generated in the collision. 
For Kerr black holes this ``hoop radius'' is~\cite{Hod:2016kzj}
\begin{equation}
 r_{\textrm{hoop}}\,=\,M+E_1+E_2+\sqrt{\left(M+E_1+E_2\right)^2-\frac{\left(J+L_1+L_2\right)^2}{\left(M+E_1+E_2\right)^2}}\,,
\label{eq:hoop_generalold}
\end{equation}
where $E_i$ and $L_i$, are the energies and angular momentum of the infalling particles. For charged particles additional terms would appear in the hoop radius, however we will restrict ourselves to consider only uncharged particles. In the following we will use the mass normalized quantities $l_ i = L_i/(M m)$, $a= \bar{a}/M$, which makes the hoop radius
\begin{align}\label{eq:hoop_general}
     r_{\textrm{hoop}} = M \left(\sqrt{\left(\frac{E_1}{M}+\frac{E_2}{M}+1\right)^2-\frac{(a+\tfrac{m}{M}  (l_1+l_2))^2}{\left(\frac{E_1}{M}+\frac{E_2}{M}+1\right)^2}}+\frac{E_1}{M}+\frac{E_2}{M}+1\right)\,.
\end{align}
All results we obtain for this setting can be extended to the collision of uncharged particles in charged black hole spacetimes.

Thus, in summary, from the outside of the black hole, only collisions at $r > r_{\textrm{hoop}}$ can, at least in principle, produce detectable signatures.

\section{Collisional Penrose process scenarios}
\label{sec:collisional_scenarios}

Having set the stage for our analysis of particle collisions, we implement the hoop conjecture and determine the center of mass energy for several scenarios at the hoop radius.

To demonstrate the algorithm we use for our analysis we start by reviewing the BSW mechanism in Sec.~\ref{sec:infinity}, for which the hoop conjecture has been already applied~\cite{Hod:2016kzj}. After that, we will present new insight about the collision between one particle that is infalling towards the black hole from infinity and one \emph{target} particle being near the black hole with almost zero (small) energy parameter $E$  and vanishing angular momentum $L$ in Sec.~\ref{sec:zero}.  In particular, we broaden the original proposal of~\cite{Hackmann:2020ogy} by extending the Schwarzschild framework to a Kerr black hole and by considering the hoop conjecture. This leads to large center of mass energy, even for non-extremal black holes, of the order of the square root of the mass of the black hole, but avoids arbitrarily large energies. Finally, we study the novel scenario of a target particle with zero energy parameter in the ergoregion, colliding with an infalling particle from infinity in Sec.~\ref{sec:zero_l}. 

\subsection{Particle collisions for infalling particles from infinity}
\label{sec:infinity}
Before discussing in details the new scenarios based on a near horizon target particle, we shall briefly review in this subsection the original scenario proposed by BSW \cite{Banados:2009pr} in which two particles coming from infinity collide near the black hole, and discuss how it is affected by the application of the hoop conjecture~\cite{Hod:2016kzj}.

In general for particles coming from infinity, their energy parameter $E$ is determined at $r \to \infty$ by the dispersion relation \eqref{eq:cas}
\begin{align}
    k_r^2 - m^2 \,=\, E^2\,.
\end{align}
 For our derivation, we assume particles which are initially at rest, i.e.~$E_1=E_2=m$.

In order to find the minimum radius for which the collision can take place, we evaluate the hoop radius~\eqref{eq:hoop_general} for the colliding particles in consideration. Introducing the ratio between the mass of the black hole and the invariant mass of the colliding particles as small parameter $1\gg \mu = m/M$, to first order in in $\mu$ the hoop radius takes the form
\begin{align}\label{eq:BSWHoopGeneral}
    r_{\text{hoop}}\,=\,M(1 + \tau + \sqrt{\mu} \rho_1 + \mu \rho_2 )\,,
\end{align}
where $\tau = \sqrt{1-a^2}$. In the non-extremal ($1>\tau>0$) and extremal case ($\tau=0$) the parameters $\rho_1$ and $\rho_2$ take the different values, namely
\begin{align}
    \rho_{1\textrm{nex}} &= 0\,,& \rho_{\textrm{2nex}} &= 2+\frac{2+2a^2-a (l_1+l_2)}{ \tau}\,,\\
    \rho_{1\textrm{ex}} &= \sqrt{2} \sqrt{4-l_1-l_2}\,,& \rho_{\textrm{2ex}} &= 2\,,
\end{align}
for which we assume that they do not become of the order $\sqrt{\mu}^{-1}$ or  $\mu^{-1}$ respectively.

Hence in the extremal case the hoop radius already acquires a correction of order $\sqrt{\mu}$, while in the non-extremal case a correction only appears at order $\mu$.

Evaluating the squared center of mass energy~\eqref{eq:cm_kerr_newman} at the general hoop radius~\eqref{eq:BSWHoopGeneral} we find to first non-vanishing (second) order in $\mu$,
\begin{align}
    \frac{E^2_{\text{cm}}(r_{\text{hoop}})}{m^2} \,=4 + \frac{l_1 (l_1-l_2)}{a^2-a l_1+(\tau +1)^2}+\frac{l_2 (l_2-l_1)}{a^2-a l_2+(\tau +1)^2}\,.
\end{align}
Clearly there is an apparent divergence when one of the angular momentum of the particles have the critical value
\begin{equation}
   l_i\,=\, \frac{a^2 + (1+\tau)^2}{a}\,.
   \label{eq:critical_angular_momentum_nonex}
\end{equation}
However, in the non-extremal case, this value cannot be reached for physical particles, since the dispersion relation is not satisfied. The situation is different in the extremal case where on finds
\begin{align}
    \frac{E^2_{\text{cm}}(r_{\text{hoop}})}{m^2} \,= \frac{2 ((l_1-4) l_1+(l_2-4) l_2+8)}{(l_1-2) (l_2-2)}\,,
\end{align}
which yields a critical angular momentum for one of the particles of $l_i=2$. Performing the calculation from the beginning, with $l_1=2$, i.e. $\rho_{1\textrm{ex}} = \sqrt{2}\sqrt{2-l_2}$, one finds
\begin{align}
    E^2_{\text{cm}}(r_{\text{hoop}})  = \,2 \left(\sqrt{2} + 1\right)\frac{m^2}{\sqrt{\mu}}\sqrt{2-l_2}\,,\label{eq:cm_kerr_limit_infinity2}
\end{align}
which is potentially very large due to the appearance of the factor $m^2/\sqrt{\mu} = m\sqrt{mM}$ but perfectly finite, as has already been pointed out in \cite{Hod:2016kzj}. The above formulas clearly show that by implementing the hoop radius one regularizes the center of mass energy, which will be otherwise infinite at the horizon, while maximizing it (given that for any radius larger than $r_{\text{hoop}}$ it will be smaller).  

\subsection{Near-horizon target particle with small energy and zero angular momentum}
\label{sec:zero}
Now we are going to consider a recently suggested scenario in which one of the particles comes from infinity, i.e., $E_2=m$, and the other particle, called the target particle, is near the horizon with small energy parameter $E_1$ and zero angular momentum $l_1=0$. In terms of the ratio between the invariant mass of the colliding particles and the black hole mass $\mu=m/M$, we assume $1\gg\mu > E_1/M$. This scenario was suggested for non-rotating black holes in \cite{Grib:2011ph,Hackmann:2020ogy} and for axially symmetric black hole spacetimes in \cite{Zaslavskii:2016uhv}. While in \cite{Hackmann:2020ogy} it was pointed out the possibility of having a squared center of mass energy arbitrarily large, we will see that when the hoop conjecture is taken into account, one finds it to be really high, but finite.

Let us first display the hoop radius again in terms of the highest order contribution in $\mu$
\begin{align}\label{eq:HoopGeneral2}
    r_{\text{hoop}}\,=\,M(1 + \tau + \sqrt{\mu} \rho_1 + \mu \rho_2 )\,,
\end{align}
where this time for the extremal ($\tau=0,a=1$) and non-extremal case
\begin{align}
    \rho_{1\textrm{nex}} &= 0\,,& \rho_{\textrm{2nex}} &= \frac{1 +a^2-a l_2}{\tau}+1\,, \label{eq_r_hoop_Kerr}\\
    \rho_{1\textrm{ex}} &= \sqrt{2} \sqrt{2- l_2}\,,& \rho_{\textrm{2ex}} &= 1 + \frac{\alpha \sqrt{2}}{\sqrt{2-l_2}} \,.\label{eq_r_hoop_Kerr2ex}
\end{align}
Again we assume that these coefficients do not become of the order $\sqrt{\mu}^{-1}$ or  $\mu^{-1}$ respectively.

The radial momentum~\eqref{eq:radial_momentum} must be real, so the term inside the square root shall be positive. Evaluating it at the hoop radius \eqref{eq:HoopGeneral2} one finds that the energy, when expanded in powers of $\mu$, must be
\begin{align}\label{eq:minimum_energy_kerr_Newman}
    E_1 \,=\alpha m \sqrt{\mu}\quad \textrm{with}\quad 
    \alpha_{\textrm{nex}}\,\geq\,\frac{\sqrt{2} \sqrt{\rho_{2\textrm{nex}}} \sqrt{\tau } (\tau +1)}{ \left(a^2+(1+\tau)^2\right)}\,,\quad \textrm{or}\quad
    \alpha_{\textrm{ex}}\,\geq \frac{\rho_{1\textrm{ex}}}{2}\,.
\end{align}
Note that $E_1$ determined here is consistent with our assumption that $\mu>E_1/M$. Using \eqref{eq:HoopGeneral2} and \eqref{eq:minimum_energy_kerr_Newman} in~\eqref{eq:cm_kerr_newman}  we find up to leading order in $\mu$ for the non-extremal and the extremal case
\begin{align}
    E^2_{\text{cm}}(r_{\text{hoop}}) \,
    &=\,\frac{m^2}{\sqrt{\mu}} 
    \tfrac{\left(a^2-a l_2+(\tau +1)^2\right) \left(\alpha_{\textrm{nex}}  (a^2+(1+\tau)^2)-\sqrt{\alpha_{\textrm{nex}}^2 \left(a^2+(\tau +1)^2\right)^2-2 \rho_{2\textrm{nex}}\ \tau  (\tau +1)^2} \right)}{\rho_{2\textrm{nex}}\ \tau  (\tau +1)^2}\,,\label{eq:cm_kerr_newman_limit_zero2}\\
    E^2_{\text{cm}}(r_{\text{hoop}}) \,
    &=\,\frac{m^2}{\sqrt{\mu}} \tfrac{2 (2 - l_2) \left(2 \alpha_{\textrm{ex}} + \sqrt{4 \alpha_{\textrm{ex}}^2-\rho_{1\textrm{ex}}^2} \right)}{ \rho_{1\textrm{ex}}^2}\,.\label{eq:cm_kerr_newman_limit_zero3}
\end{align}
As for the extremal BSW scenario discussed in the previous section, we find that at the hoop radius the center of mass energy in a collision between a target particle of small energy and zero angular momentum and an infalling particle from infinity is finite, albeit potentially very large, since its leading order term is proportional to $m^2/\sqrt{\mu} = m \sqrt{mM}$, which involves the black hole mass $M$. Most interestingly, and in contrast to the original BSW scenario, for the target particle scenario this already happens for non-extremal black holes.

\subsubsection{The hoop conjecture in Schwarzschild geometry}
In~\cite{Grib:2011ph,Hackmann:2020ogy} the emergence of an arbitrarily large center of mass energy was discussed for a non-rotating and non-charged black hole. In this part of the article we explicit show that this energy can be really large, of the order of $ m \sqrt{mM}$, as previously displayed, but not arbitrarily large. We are able to describe this particular scenario by setting $a=0$ and $\tau=1$ in the previously displayed equations.

The hoop radius of a non-rotating Schwarzschild black hole in the presence of a target and incoming particle is 
\begin{equation}
    r_{\textrm{hoop}} \,=\, 2 (M+m)\,.
\end{equation}
The energy of the target particle follows from \eqref{eq:minimum_energy_kerr_Newman}
\begin{equation}\label{eq:minimum_energy_schw}
    E_1 \,=\, \alpha m \sqrt{\mu}\,,\qquad \text{with}\qquad \alpha\,\geq\, 1\,,
\end{equation}
and the squared center of mass energy can be read off from \eqref{eq:cm_kerr_newman_limit_zero2}
\begin{equation}
    E^2_{\text{cm}}(r_{\text{hoop}})\,=\, \frac{m^2}{\sqrt{\mu}}2(\alpha - 2 \left(\sqrt{\alpha ^2-1} \right))\,.
    \label{eq:cm_small2}
\end{equation}
Thus, by taking the hoop conjecture into account, the squared center of mass energy cannot be arbitrarily large. It is again proportional to $m\sqrt{ m M}$, which is finite. The mass of the black hole determines the scale of the center of mass energy of the collision.

In the spherically symmetric case another limiting factor is the question on the existence of target particles. It has been pointed out that these may not exist for uncharged spherically symmetric black holes \cite{Zaslavskii:2020agl}.

\subsection{Target particle with zero energy and small angular momentum}\label{sec:zero_l}
Finally, we propose a new scenario based on the fact that for rotating black holes it is possible that the target particle in the ergosphere, which has zero energy $E_1=0$ (see for example~\cite{Zaslavskii:2016uhv,Grib:2016mak}) and small angular momentum $l_1 \sim \gamma \sqrt{\mu}$. Indeed, in this case, a diverging center mass energy at the horizon according to \eqref{eq:COMHor} could in principle be obtained.

This scenario can be treated analogously to the one for particles with small energy and zero angular momentum. We identify the leading order in $\mu$ coefficients in the hoop radius \eqref{eq:HoopGeneral2} to be
\begin{align}
    \rho_{1\textrm{nex}} &= 0\,,& \rho_{\textrm{2nex}} &= \frac{1 +a^2-a l_2}{\tau}+1\,,\\
    \rho_{1\textrm{ex}} &= \sqrt{2} \sqrt{2- l_2}\,,& \rho_{\textrm{2ex}} &= \frac{\gamma }{\sqrt{2}\sqrt{2- l_2}}+1\,.
\end{align}

The positivity requirement on the radial momentum \eqref{eq:radial_momentum} fixes a small angular momentum $l_1$ (instead of the energy $E_1$ in the previous section) to be of the form
\begin{equation}
l_1\,=\,-\frac{\gamma \sqrt{m}}{ \sqrt{M}}\,,\qquad \text{with}\qquad \gamma_{\textrm{2nex}}\,\geq\, \frac{\sqrt{2} \sqrt{\rho_{\textrm{2nex}}} \sqrt{\tau } (\tau +1)}{a}\,, \qquad or \qquad \gamma_{\textrm{ex}}\,\geq\, \rho_{\textrm{1ex}}\,.
\label{eq:minimum_energy_kerr_Newman_l}
\end{equation}
Noticeably, this quantity seems to diverge when $a=0$, but this is just the manifestation that zero-energy particles do not exist in the non-rotating case. Combining these findings in the squared center of mass energy we obtain for the exremal and non-extremal case
\begin{align}
    E^2_{\text{cm}}(r_{\text{hoop}}) \,
    &=\,\frac{m^2}{\sqrt{\mu}} \frac{\left(a^2-a l_2+(\tau +1)^2\right) \left(a \gamma -\sqrt{a^2 \gamma ^2-2 \rho_{\textrm{2nex}} \tau  (\tau +1)^2}\right)}{\rho_{\textrm{2nex}} \tau  (\tau +1)^2} \,,\label{eq:cm_kerr_newman_limit_zero2_l}\\
    E^2_{\text{cm}}(r_{\text{hoop}}) \,
    &= \frac{m^2}{\sqrt{\mu}} \frac{2 (2 - l_2) \left(\sqrt{\gamma ^2-\rho_{1\textrm{ex}}^2}+\gamma \right)}{\sqrt{\mu } \rho_{1\textrm{ex}}^2} \, \,.\label{eq:cm_kerr_newman_limit_zero3lsmall}
\end{align}
As in the small $E_1$ and $l_1=0$ case, for collisions at the hoop radius the center of mass energy stays finite and is of order $m\sqrt{ m M}$. In contrast, when the hoop conjecture is not applied, one obtains the possibility of a diverging center of mass energy, see \cite{Grib:2010dz}.

\section{Astrophysical black holes as particle accelerators?}
\label{sec:stellarBHs}

So far, we derived the center of mass energy of different types of particle collisions in Kerr spacetimes, by taking the hoop conjecture into account. We found that the hoop conjecture avoids infinite center of mass energies for all kinds of collisions: between infalling particles from infinity and between one infalling and one target particle.

Nevertheless, the hoop conjecture does not imply that the center of mass energies in particle collisions must be necessarily small, their value can still be large. Generically we found scenarios for which it is of order $m\sqrt{mM}$, even for non-extremal black holes.  The question now is, if particles with high energy, which are produced in such collisions,  can escape all the way up to infinity, rather than only interact with other environmental particles before plunging into the black hole.

To this end, we shall consider in what follows a $2-2$ interaction and compute the energy of the emitted particles to get an insight about how such high center of mass energy collisions would manifest itself in observations far away from the black hole. We obtain that the energy of a massless particle which is produced in such a collision and could escape the ergosphere with an energy of the same order of the one of the particle coming from infinity,  in agreement with previous analyses for the BSW scenario~\cite{Bejger:2012yb,Leiderschneider:2015kwa}. 

\subsection{Dependence on the angular momentum paramater}
Besides the hoop conjecture, one main argument against high center of mass collisions~\cite{Berti:2009bk,Jacobson:2009zg,Gao:2011sv} is that black holes cannot become extremal by accretion~\cite{Thorne:1974ve}. Theoretical consideration limit their maximal angular momentum in astrophysical setting to $a=0.998$. However, note that black holes which are created in black hole collision processes this limit can be exceeded \cite{Kesden:2008ga}, and also in super-Eddington accretion processes such limit has been claimed to be possibly violated~\cite{Sadowski:2011ka}. For these reason, we shall not limit our analysis here to values $a\leq 0.998$. 

In contrast to the standard BSW scenario, as we pointed out in Secs.~\ref{sec:zero} and \ref{sec:zero_l}, both target particle scenarios lead to a very high center of mass energies even for non extremal uncharged rotating black holes. Actually, the target particle scenario discussed in Sec.~\ref{sec:zero} does so even for non-rotating black holes~\cite{Hackmann:2020ogy}. Let us evaluate the center of mass energies \eqref{eq:cm_kerr_newman_limit_zero2} and \eqref{eq:cm_kerr_newman_limit_zero3} for the values $l_2=0$ and $\alpha=1.1$ for the several angular momentum parameters $a$:
\begin{equation}
\begin{split}
    E^2_{\text{cm}}(r_{\text{hoop}}) \,&=\, 1.3\, m\sqrt{m M}\qquad \text{for}\qquad a\,=\,0.3\,,\\
    E^2_{\text{cm}}(r_{\text{hoop}}) \,&=\, 1.3\, m\sqrt{m M}\qquad \text{for}\qquad a\,=\,0.998\,,\\
     E^2_{\text{cm}}(r_{\text{hoop}}) \,&=\, 3.1\, m\sqrt{m M}\qquad \text{for}\qquad a\,=\,1\,.
  \end{split}
\label{eq:cm_kerr_newman_limit_zero_lexamples}
\end{equation}
It is clear from above that the center of mass energy is weekly sensitive on the black hole spin.

To get an insight of how high the center of mass energies can become,  we consider a collision of neutrons of $m = 939.6 \times 10^6 \text{eV}$ and a black holes spin of $a=0.998$. We start by giving a numerical estimation of the hoop radius. For the mass and the spin considered above, and for a mass of a stellar black holes of $10$ solar masses $(M =1.12  \times 10^{67} \text{eV})$,  the hoop radius computed from~\eqref{eq:HoopGeneral2} is very small, viz
\begin{equation}
\frac{r_{\text{hoop}}-r_{\text{hor}}}{r_{\text{hor}}}\,=\, 2.6\times 10^{-57}\,,
\end{equation}
where $r_{\text{hor}}$ is the radius of the horizon. Even if the hoop radius is going to be placed just outside the horizon by a tiny bit, it is still relevant because it provides the minimal radius (at which the reaction can happen without being engorged by a new horizon), while maximizing the center of mass energy of its outcome and make it indeed finite. Placing the collision at an arbitrary distance from the horizon we only reduce the center of mass energy, while placing it below the hoop radius would most probably lead to the formation of a new horizon and the entrapment of the particles. 

Now we compute the center of mass energy  for two types of black holes:
\begin{itemize}
    \item stellar black holes of $10$ solar masses $(M =1.12  \times 10^{67} \text{eV})$ yield energies of the order
    \begin{equation}
        E_{\text{cm}}(r_{\text{hoop}}) \,=\,5.5 \times 10^{23}\,\text{eV}\,,
    \label{eq:cm_kerr_newman_limit_zero_lexamples2}
    \end{equation}
    \item the most massive black holes which have been detected with a mass of $10^{10}$ solar masses~\cite{Shemmer:2004ph,Mehrgan_2019} would allow for energies of the order
    \begin{equation}
        E_{\text{cm}}(r_{\text{hoop}}) \,=\,9.7 \times 10^{25}\textrm{eV}\,.
    \label{eq:cm_massiveBH}
    \end{equation}
\end{itemize}
Clearly, these energies do not reach the Planck energy $E_{\text{Pl}} = 1.22\times 10^{28}\, \textrm{eV}$, but stay three orders of magnitude below that. Nevertheless, the energies which can be produced by a ``target particle-infalling particle'' collisional Penrose process in the ergosphere of a rotating non-extremal black hole are so large that they reach the energies of the order of those detected in ultra-high-energy cosmic rays (UHECR)~\cite{Ivanov:20198M,Aab:2020gxe,Deligny:2020gzq}. However, as we will see in the following, the energies of the particles actually able to escape to infinity do not reach these large values. 

\subsection{On the fate of the produced particles}
\label{sec:collision}
In~\cite{Bejger:2012yb,Leiderschneider:2015kwa} the products of the collision in the BSW scenario have been considered. In particular, it has been shown that, in a $2-2$ collision, the maximum energy of the particles reaching infinity is just $1.3$ or $6$ times the sum of the mass of the infalling particles. Here we are going to consider (following the same steps) the scenario of Sec.~\ref{sec:zero}, in which the target particle has zero angular momentum and very small energy (analogous results can be obtained for the scenario discussed in Sec.~\ref{sec:zero_l}). Notice that in~\cite{Bejger:2012yb,Leiderschneider:2015kwa} the hoop conjecture was not considered, while here it is a central point in our analysis.

\subsubsection{Escape conditions}\label{ssec:esccond}

As shown above, the novel processes discussed in Secs.~\ref{sec:zero} and~\ref{sec:zero_l} can achieve very large center of mass energies only for collisions very close to the horizon, albeit they cannot be arbitrary close given the hoop conjecture, $r_\text{minimal}\geq r_\text{hoop}$. In this part of the paper we discuss the conditions under which particles produced at the hoop radius can escape to infinity, before we discuss next the possible energies attainable by these particles. 
Following the analysis of~\cite{Bardeen:1972fi,chandrasekhar1998mathematical,Bejger:2012yb,Leiderschneider:2015kwa}, where a 2--2 collision is considered, one can 
see that one of the produced pair particles (we consider it as particle 3) can escape to infinity in two different cases:
\begin{itemize}
    \item The  newly produced pair consists of an ingoing positive energy particle and an outgoing negative energy one (which will hence be confined in the ergoregion). The positive energy, ingoing particle, can escape only by bumping off the Kerr black hole effective potential, turn into an outgoing one and then reach infinity. 
    
    In this case, a standard analysis of Kerr black hole orbits shows that the ingoing particle can escape if it is produced at a radius not smaller than a critical radius~\cite{Leiderschneider:2015kwa}
    \begin{equation}
        r_u\,=\, 2 M\left(1+ \cos\left[\frac{2}{3} \arccos(-a)\right]\right)    \,,
    \label{eq:ru}
    \end{equation}
and is characterized by an impact parameter, $b=L/(EM)$, greater than the critical value
\begin{equation}
    b_u\,=\,-6 \cos\left[(2\pi+\arccos{a}/3)\right]-a\,.
    \label{eq:bu}
\end{equation}
Given that $r_u\geq r_+$ always in these coordinates~\cite{Bardeen:1972fi},  where $r_+$ is the outer horizon, in general, we have for generic angular momentum parameter $a$ that $r_u>r_\text{hoop}$ given that $r_\text{hoop}$ is close to but larger than $r_+$. Hence, the collision would have to happen at $r_u$ rather than $r_\text{hoop}$ for the particles to escape to infinity with the consequence that the center of mass energy would be not as high as we derived in \eqref{eq:cm_kerr_newman_limit_zero2} and \eqref{eq:cm_kerr_newman_limit_zero2_l}.

However, as the black hole gets closer to extremality, one can show that $r_u\to r_\text{hoop}$ and that there are special situations such that $r_\text{hoop}\geq r_u$ for a particular range of the parameter~$a$. The latter is precisely the situation needed so a particle produced in a collision at $r_\text{hoop}$ can escape to infinity.

In order to see that this can really happen, let us take $a=1-\epsilon$ and perform a power series expansion of $r_u$ in~$\epsilon$
\begin{equation}
r_u\,\approx\, M\left(1+2 \sqrt{\frac{2}{3}\epsilon}\right)   \,.
    \label{eq:ru_epsilon}
\end{equation}
Doing the same for the hoop radius defined by the expressions in Eq.~\eqref{eq_r_hoop_Kerr} (neglecting terms just of order $m$, $\epsilon$, and $m\sqrt{\epsilon}$), we find
\begin{equation}
r_\text{hoop}\,\approx\, M+ \frac{2 -l_2}{\sqrt{2\epsilon}}m+ \sqrt{2 \epsilon} M   \,.
    \label{eq:rhoop_epsilon}
\end{equation}
Therefore, the condition $r_\text{hoop}\geq r_u$ implies
\begin{equation}
\frac{m(3+2\sqrt{3})(2-l_2)}{2 M}\,\geq\,\epsilon \,>\, 0  \,.
    \label{eq:epsilon_r1}
\end{equation}
For the extremal case, $\epsilon=0$, we found a separate expression of the hoop radius, so we have to start from Eq.~\eqref{eq_r_hoop_Kerr2ex}, which obviously is independent of $\epsilon$, and always greater than $r_u$, since, in this case $r_u=r_+=M$ and, as mentioned before, $r_\text{hoop}$ is always greater than $r_+$. Thus, in the above formula we can replace the last strictly greater by~$\epsilon\geq 0$.

The above upper bound value of $\epsilon$ is indeed very small but non-zero, showing that for near-extremal Kerr black holes the high-energy particles produced at the hoop radius described in Secs.~\ref{sec:zero} and~\ref{sec:zero_l}  can potentially escape to infinity. As said, the additional condition to be satisfied for this to happen is that the impact parameter of the produced infalling particle must satisfies $b_3\geq b_u$.

Making a power series expansion in Eq.~\eqref{eq:bu} around $\epsilon$ for $a=1-\epsilon$ we find
\begin{equation}\label{eq:b_epsilon}
b_u\,=\,2+\sqrt{6 \epsilon}\,,
\end{equation}
Now, using the maximum value of $\epsilon$ in Eq.~\eqref{eq:epsilon_r1}, for which $r_\text{hoop}=r_u$, we find from the previous expression that
\begin{equation}
b_u\,=\,2+\sqrt{3(3+2\sqrt{3})(2-l_2)} \sqrt{\frac{m} {M}}\,,
    \label{eq:b_min}
\end{equation}
which is basically equal to 2 for the values of $m$ and $M$ considered in the previous section. This implies that for any impact parameter $b_3$ equal or greater than this value, initially ingoing particles can escape to infinity after bouncing off the potential. 


\item In the other possible scenario, the newly produced pair consists of an ingoing negative energy particle, and an outgoing positive energy one. The positive energy, outgoing  particle, can reach infinity if its impact parameter is smaller than $b_u$ and greater than a minimum value $b_l$ (which in the extremal case is $b_l=-7$, while $b_u=2$~\cite{Leiderschneider:2015kwa}). Note that in this case, it is not necessary for $r_\text{hoop}$ to be greater than $r_u$, as any outgoing particle with impact parameter in the above range will never meet a turning point on its orbit.

Now, $b_u$ is minimal for $a=1$ ($b_u=2$), and increases for decreasing $a$ (see for example Eq.~\eqref{eq:bu}).  Consequently, it is enough to ask that our ingoing particle has an impact parameter $b_3$ less than $2$.  

\end{itemize}

Thus, in summary, we just saw that for nearly extremal black holes, i.e., black holes with $a=1-\epsilon$, where $\epsilon$ is very small, particles produced at the hoop radius can escape to infinity. In the following, we derive the maximal energy of such escaping photon. 

\subsubsection{Maximum energy of the emitted particles}
\label{sec:maximum_energy}
In order to study the maximum energy of the emitted particles, we need to impose the conservation laws of four-momenta between the initial ($1,2$) and final ($3,4$) particles
\begin{equation}
     k_{1\mu}+k_{2\mu}\,=\, k_{3\mu}+k_{4\mu}\,.
\end{equation}
As in~\cite{Bejger:2012yb,Leiderschneider:2015kwa}, we consider that the emitted particles are photons.
Then, we can easily solve the previous equation by considering its norm  
\begin{align}
    0 \,=\, g^{\mu\nu}(k_{1\mu}+k_{2\mu} - k_{3\mu})(k_{1\nu}+k_{2\nu} - k_{3\nu}) = -E^2_{CM} - 2 g^{\mu\nu}(k_{1\nu}+k_{2\nu}) k_{3\mu} \,,
    \label{eq:e3}
\end{align}
where we have used Eq.~\eqref{eq:cm} and that $g^{\mu\nu}k_{\mu3}k_{\nu3}=g^{\mu\nu}k_{\mu4}k_{\nu4}=0$ for photons. It turns out that \eqref{eq:e3} becomes a linear equation for $E_3$ and thus
\begin{align}\label{eq:E31}
    E_3 \,=\, \frac{E^2_{CM}}{\mathfrak{a} }  \,,
\end{align}
where $\mathfrak{a}$ is obtained from the second term in \eqref{eq:e3}. It multiplies $E_3$ and is a lengthy expression of the metric components.

Studying the non-extremal target particle scenario, as discussed in Section \ref{sec:zero}, in which the colliding particles have parameters $E_2=m$, $E_1=\alpha m \sqrt{\mu}$ and $l_1=0$, hoop radius defined by~\eqref{eq_r_hoop_Kerr} and center of mass energy given in \eqref{eq:cm_kerr_newman_limit_zero2}, we find the leading orders of magnitude of these terms are
\begin{align}\label{eq:coefforders}
        E^2_{CM} \sim M^2 \sqrt{\mu}^3 = m \sqrt{m M} \,,\quad \mathfrak{a} \sim M \sqrt{\mu} = \sqrt{m M} \,.
\end{align}
To leading order in $\mu$ the energy of the escaping particle is thus given by
\begin{align}\label{eq:energy3}
        E_3{}_{\textrm{nex}} = m \tfrac{\left(a^2-a l_2 + (\tau +1)^2\right) \left(\alpha_{\textrm{nex}}  \left(a^2+(\tau +1)^2\right)-\sqrt{\alpha_{\textrm{nex}} ^2 \left(a^2+(\tau +1)^2\right)^2-2 \rho_{\textrm{2nex}} \tau  (\tau +1)^2}\right)}{\left(a^2-a b_3+(\tau +1)^2\right) \left(\alpha_{\textrm{nex}}  \left(a^2+(\tau +1)^2\right)-\sigma  \sqrt{\alpha_{\textrm{nex}} ^2 \left(a^2+(\tau +1)^2\right)^2-2 \rho_{\textrm{2nex}} \tau  (\tau +1)^2}\right)}\,,
\end{align}
where $\sigma = 1$ for a particle that is infalling after its production and then bounces of the potential to reach infinity, while $\sigma=-1$ for a particle that is directly outgoing and escaping to  infinity. Clearly both scenarios posses a critical value of the impact parameter such that the denominator vanishes
\begin{equation}
    b_3\,=\, \frac{2(1+\sqrt{1-a^2})}{a}\,.
\label{eq:b_horizon}
\end{equation}

Now, investigating the behaviour of $E_3{}_{\textrm{nex}}$ for nearly extremal black holes $a=1-\epsilon$ and $b_3 = 2 + \sqrt{6\epsilon}$, as found in the previous section, we obtain, using the maximal value of $\epsilon = X \mu$  as displayed in equation \eqref{eq:epsilon_r1} ,
\begin{align}
\label{ENearEx}
    E_3{}_{\textrm{nearly-ex}} = m 
    \tfrac
    {\left(2 \alpha +\sqrt{2} \sqrt{2 \alpha ^2+l_2-2}\right)}
    {\left(\sqrt{\left(14 \sqrt{3}-4 \sqrt{48 \sqrt{3}+81}+27\right)} \sigma +\left(2 \sqrt{2 \sqrt{3}+5}-\sqrt{6 \sqrt{3}+9}\right)\right) \sqrt{2-l_2}}
\end{align}
which is of order $m$. 
However, compared to the BSW scenario considered in~\cite{Bejger:2012yb,Leiderschneider:2015kwa} one can find some amplification: for example taking $l_2 = 2 - 10^{-4}$ (i.e. close but not infinitesimally close to $2$) one is able to produce an escaping particle with an energy one hundred times greater than the incoming particle.
The limiting value $l_2=2$ appears to be critical as a divergence seems to arise in the above expression.\footnote{A superficial inspection of Eq.~\eqref{ENearEx} might induce to think that very high energy could be achieved even before reaching actual extremality, i.e.~for $l_2\to 2$. This is however not true as the found expression is already valid near extremality with $\epsilon = X \mu$, considering very small $X$ would require a next order set of expansion that would end up modifying the expression for $E_3{}_{\textrm{nearly-ex}}$ so to confirm the smallness of the attainable energy for the escaping particle.} However, this case needs to be treated separately given that it is associated to an extremal black hole for which $\epsilon=0$.

Let us then also consider the extremal black hole case for the target particle scenario. The hoop radius and center of mass energy are modified and provided respectively by Eq.~\eqref{eq_r_hoop_Kerr2ex} and Eq.~\eqref{eq:cm_kerr_newman_limit_zero3}. The order of $\mathfrak{a}$ is different for the infalling ($\sigma=1$) and the outgoing ($\sigma=-1$) produced particle. For the infalling and bouncing particle we find
\begin{align}
    E_3{}_{\textrm{exIn}} = m\frac{(l_2-2)}{b_3-2}\,,
\end{align}
while for the directly escaping particle we have
\begin{align}
     E_3{}_{\textrm{exOut}} = m \sqrt{\frac{m}{M}} \frac{\left(2 \alpha_{\textrm{nex}} + \sqrt{4 \alpha_{\textrm{ex}}^2-\rho_{\textrm{1ex}}^2}\right)}{2 (2-b_3)}\,.
\end{align}
As in the nonextremal case we see a possible divergence for $E_3$, this time for $b_3=2$. However, when one investigates the behaviour of $E_3$ for $b_3=2$ we find for ingoing and outgoing particle
\begin{align}
    E_3{}_{\textrm{ex}} = m \frac{\left(2 \alpha_{\textrm{nex}} + \sqrt{4 \alpha_{\textrm{ex}}^2-\rho_{\textrm{1ex}}^2}\right)}{\left( 2 + \sigma \sqrt{3} \right) \rho_{\textrm{1ex}}}\,.
\end{align}

Therefore, even though the center of mass energy of the collision is large, of order $m\sqrt{mM}$, the energy of the escaping particles in the target particle scenarios stays small, of order $m$. This is compatible, with what was found in the literature for the BSW scenario~\cite{Bejger:2012yb,Leiderschneider:2015kwa}.


\section{Conclusion}
\label{sec:conclusions}

We have studied and extended the analysis of the collisional Penrose process based on the target particle scenario. Such scenarios were claimed to lead to arbitrary high center of mass energies even for non-rotating black holes. In particular, we have considered Kerr black holes and added the physical requirement that the hoop conjecture is not violated in the collisional process. 
We found that, when the hoop conjecture is applied, infinite center of mass energy collisions are avoided outside of the event horizon of any rotating (or non-rotating) black hole. This is due to the fact that the divergences appear only at the horizon. By considering here  the hoop conjecture, we use the closest distance to the horizon for which the particle does not have to be swallowed by the black hole, and therefore, the energies involved are maximum.
Nonetheless, for the target particle scenario, the squared center of mass energy is always of order $m\sqrt{m M}$, where $m$ is the mass of the particles and $M$ is the mass of the black hole, even for non-extremal black holes. Thus for very heavy black holes, large center of mass energies are possible. The most massive black holes observed, $10^{10}$ solar masses, lead to energies of order $10^{25}\, \textrm{eV}$, slightly higher than the highest energies detected in UHECRs.



However, we have also studied the energy of the primary products produced in such collisions (i.e.~in the target-particle scenario) and found that while their energies can indeed be very high, only those particles with energies of the same order of the parent particle coming from infinity are able to escape to infinity. This implies that such particles will have relatively low energies. This result agrees with the one obtained in~\cite{Bejger:2012yb,Leiderschneider:2015kwa} for the BSW scenario, and it seems to suggest a general limitation of any collisional Penrose process concerning the energy of the primary products able to reach infinity. Nonetheless, one cannot at the moment exclude that secondary products --- generated via repeated near horizon collisions of the primary particles with particles in astrophysical environments --- could lead to high energy fluxes at infinity. This is an open question that we hope this study will stimulate to further investigate in the next future.


\acknowledgments

SL acknowledges funding from the Italian Ministry of Education and Scientific Research (MIUR) under the grant PRIN MIUR 2017-MB8AEZ. CP was funded by the Deutsche Forschungsgemeinschaft (DFG, German Research Foundation) - Project Number 420243324. JJR acknowledges support from the INFN Iniziativa Specifica GeoSymQFT. The authors would like to acknowledge networking support by the COST Action QGMM (CA18108). We like to thank the unknown referees for very useful discussions and comments that improved this work.


\bibliography{GR}
\bibliographystyle{JHEP}

\end{document}